\documentclass[prl,twocolumn,showkeys,showpacs,superscriptaddress,floatfix,longbibliography]{revtex4-1}

\usepackage[T1]{fontenc}
\usepackage{epsfig,psfrag}
\usepackage{latexsym}
\usepackage{graphicx}
\usepackage{dcolumn}
\usepackage{amssymb}
\usepackage{amsmath}
\usepackage{bm}
\usepackage{mathrsfs} 
\usepackage{bigints}
\usepackage{upgreek} 
\usepackage{color}
\usepackage{longtable}
\usepackage{xspace} 
\usepackage{siunitx}
\usepackage{epstopdf}
\usepackage{float}  
\usepackage[normalem]{ulem} 
\usepackage{hyperref}
\hypersetup{
    colorlinks=true,
    citecolor=blue,
    linkcolor=blue,
    filecolor=blue,      
    urlcolor=blue,
    pdftitle={Superfluid-Mott transition in a frustrated triangular optical lattice}, 
    }
\usepackage{bbold} 
\usepackage{bbm}
\usepackage{placeins}
\usepackage[margin=2cm]{geometry}

\definecolor{mygreen}{RGB}{20,148,20}
\definecolor{myviolet}{RGB}{127,0,255}

\usepackage{filemod}

\newcommand{\frs}{frustrated } 
\newcommand{\ufrs}{unfrustrated }

\newcommand{\unt}[1] {\,{\mathrm{#1}}} 

\usepackage{physics}
\usepackage[version=4]{mhchem}

\usepackage{soul}

\DeclareSIUnit\azero{$\ensuremath{a_0}$}
\DeclareSIUnit\Er{$\ensuremath{E_\mathrm{R}}$}

\begin{document}
	
    \title{Superfluid-Mott transition in a frustrated triangular optical lattice}
    
    \author{Mehedi Hasan}
    \email{m.hasan1@imperial.ac.uk}
    \affiliation{Cavendish Laboratory, University of Cambridge, J.J. Thomson Avenue, Cambridge CB3 0US, UK}
    \affiliation{Centre for Cold Matter, Blackett Laboratory, Imperial College London, London SW7 2AZ, UK} 

    \author{Luca Donini}
    \affiliation{Cavendish Laboratory, University of Cambridge, J.J. Thomson Avenue, Cambridge CB3 0US, UK}
    
    \author{Sompob Shanokprasith}
    \affiliation{Cavendish Laboratory, University of Cambridge, J.J. Thomson Avenue, Cambridge CB3 0US, UK} 
    
    \author{Daniel Braund}
    \affiliation{Cavendish Laboratory, University of Cambridge, J.J. Thomson Avenue, Cambridge CB3 0US, UK}
    
    \author{Tobias Marozsak}
    \affiliation{Cavendish Laboratory, University of Cambridge, J.J. Thomson Avenue, Cambridge CB3 0US, UK}
    
    \author{Moritz Epping}
    \affiliation{Cavendish Laboratory, University of Cambridge, J.J. Thomson Avenue, Cambridge CB3 0US, UK}
    
    \author{Daniel Reed}
    \affiliation{Cavendish Laboratory, University of Cambridge, J.J. Thomson Avenue, Cambridge CB3 0US, UK}
    \affiliation{QuEra Computing UK, Harwell Science and Innovation Campus, OX11 0GD, UK}
    
    \author{Max Melchner}
    \affiliation{Cavendish Laboratory, University of Cambridge, J.J. Thomson Avenue, Cambridge CB3 0US, UK}
    \affiliation{Max-Planck-Institut für Quantenoptik, 85748 Garching, Germany}
    \affiliation{Fakultät für Physik, Ludwig-Maximilians-Universität, 80799 Munich, Germany}
    
    \author{Tiffany Harte}
    \affiliation{Cavendish Laboratory, University of Cambridge, J.J. Thomson Avenue, Cambridge CB3 0US, UK}
    
    \author{Ulrich Schneider}
    \email{uws20@cam.ac.uk}
    \affiliation{Cavendish Laboratory, University of Cambridge, J.J. Thomson Avenue, Cambridge CB3 0US, UK}

\begin{abstract}
Geometric frustration can significantly increase the complexity and richness of many-body physics and, for instance, suppress antiferromagnetic order in quantum magnets.
Here, we employ ultracold bosonic $^{39}$K atoms in a triangular optical lattice to study geometric frustration by stabilizing the gas at the frustrated upper band edge using negative absolute temperatures.
We find that geometric frustration suppresses the critical interaction strength for the (chiral-)superfluid to Mott insulator ($\chi$-SF--MI) quantum phase transition by a factor of 2.7(3) and furthermore changes the critical dynamics of the transition. Although the emergence of coherence during fast ramps from MI to the ($\chi$-)SF regime is continuous in both cases, for ramps longer than a few tunnelling times, significant differences emerge. In the \frs case, no long-range order emerges on the studied timescales, highlighting a significantly reduced rate or even saturation of the emerging coherence. 
This work opens the door to quantum simulations of frustrated systems that are often intractable by classical simulations.

\end{abstract}
\maketitle 

	\section{Introduction} 
Ultracold atoms in optical lattices offer a versatile, highly controlled, and pristine platform for the quantum simulation of strongly-correlated matter. A paradigmatic example is bosonic atoms, where the ground state is typically either a long-range ordered superfluid (SF) or an unordered Mott insulator (MI), depending on whether the kinetic energy or the on-site repulsion dominates~\cite{Jaksch_PRL_1998, Greiner2002}. The realization of a wide range of geometries including cubic, square, triangular, honeycomb, Lieb, and kagome lattices~\cite{Jo2012, Braun2013, Taie2015, Li2016, Trenkwalder2016,  Yang2021, Song2022, Brown2022, Xu2023}, coupled with the precise control of interactions via Feshbach resonances, enables the exploration of both novel phases and quantum phase transitions as well as their out-of-equilibrium dynamics.

The triangular lattice is the simplest example of a geometrically-frustrated lattice where the lattice structure itself can suppress the emergence of long-range order. The clearest examples of this are magnetic systems~\cite{Vojta2018,Diep2020}. With ferromagnetic interactions, Ising spins on a triangular lattice show long-range order at low temperatures. For antiferromagnetic interactions, however, the situation is rather different and even the ground state is highly degenerate and does not show long-range order~\cite{Wannier1950}. Geometric frustration can significantly increase the complexity and richness of the emerging many-body physics and gives rise to a multitude of phenomena ranging from the finite configuration entropy of water~\cite{Pauling1935} to spin ice~\cite{Skjrv2019} or spin liquids~\cite{Anderson1973}.
In particular, it can render seemingly simple models effectively intractable by classical simulations. As a prominent example, the antiferromagnetic Heisenberg model on the kagome lattice is often discussed as a candidate for spin-liquid states~\cite{Anderson1973, Yan2011, Zhou2017}. However, the precise nature of the ground state remains unresolved despite significant numerical efforts~\cite{Luchli2019}.

\begin{figure*}[htb] 
    \centering
    \includegraphics[width=0.98\textwidth]{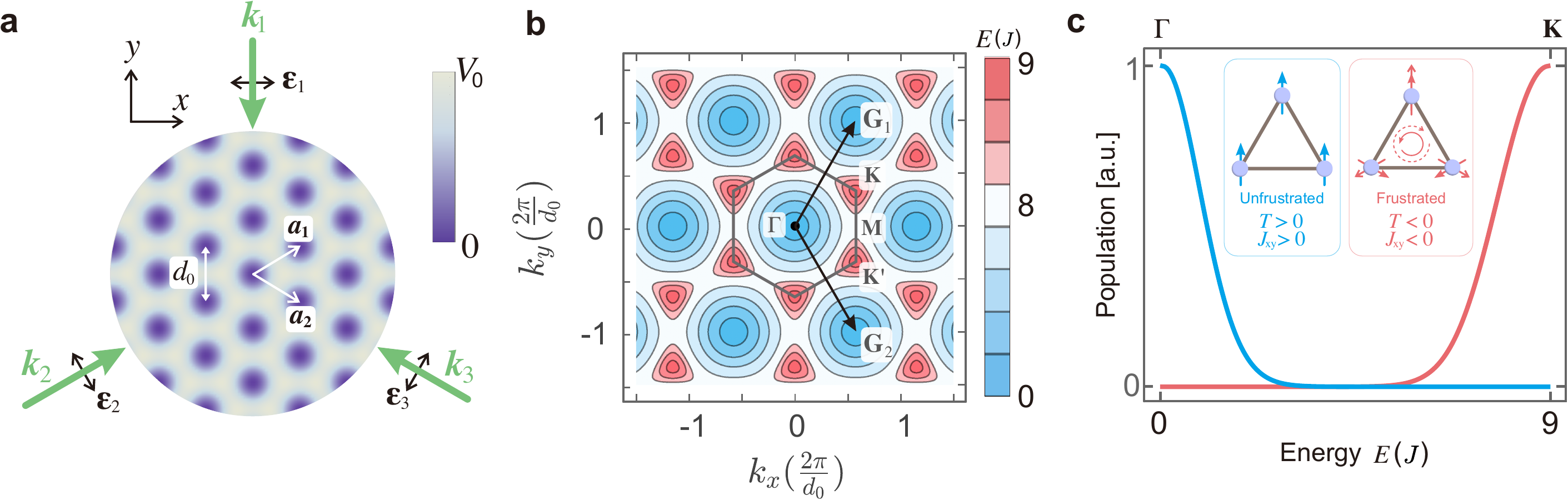}  
    \caption{ \textbf{Triangular lattice, band structure, and frustration.} \textbf{a}, Three interfering blue-detuned laser beams create a triangular lattice with lattice spacing $d_0$ and lattice vectors $\mathbf{a}_{1,2}$; the wave vectors $\mathbf{k}_{1,2,3}$ of the beams are shown by the green arrows while the black arrows indicate the in-plane polarizations $\mathbf{\varepsilon}_{1,2,3}$. In this blue-detuned lattice, the atoms sit around the zero-intensity points (purple regions). \textbf{b}, Tight-binding dispersion relation of the lowest band, with the hexagon outlining the first Brillouin zone. 
    The inherent geometric frustration of the triangular lattice leads to two non-equivalent degenerate maxima at the $\mathbf{K}$ and $\mathbf{K'}$ points.
    \textbf{c}, Equilibrium population of an ideal Bose gas at positive ($T = 2|J|$, blue) and negative ($T = -2|J|$, red) absolute temperatures.
    For $T>0, T\rightarrow0$  the bosons condense at the minimum of the band at the $\mathbf{\Gamma}$ point, unaffected by the geometric frustration.  For $T<0$, however, condensation occurs at the maxima of the band, namely the $\mathbf{K}$ or $\mathbf{K'}$ point. Inset: For $J_\mathrm{XY}>0$, the system resembles a ferromagnetic spin configuration with all spins  (blue arrows) being aligned.  However, for an antiferromagnetic system ($J_\mathrm{XY}<0$) the frustration is evident, leading to two degenerate spin-configurations with opposite chirality, shown schematically by two counter-rotating arrows. } \label{fig:fig1} 
\end{figure*}

Frustration plays a central role also in the antiferromagnetic Heisenberg model on the triangular lattice: Similarly to the Ising case, antiferromagnetic ordering is unfavourable along the out-of-plane direction such that long-range N\'{e}el order emerges with all spins lying in the lattice plane~\cite{Capriotti1999, White2007}. These states form $120^{\circ}$-angles between neighbouring spins and, as a consequence, carry an intrinsic chirality, where the spins around a plaquette rotate either clockwise or counter-clockwise~\cite{Diep2020}, shown schematically by the two counter-rotating arrows in Fig.~\ref{fig:fig1}\textbf{c}. 
The transition into the ordered state thus breaks two symmetries, the $\mathrm{U}(1)$ symmetry related to the in-plane N\'{e}el order and the Ising-type $\mathbb{Z}_2$ symmetry connected with the two possible chiralities. This leads to several distinct possibilities for the nature of the transition as the two symmetries can be broken either concurrently~\cite{Plakhty2000, Yosephin1985} or sequentially~\cite{Quirion2006, Takeda1986}. In the former case, the transition can be discontinuous (i.e., first order) or continuous, while a sequential breaking suggests two continuous phase transitions. This has been a matter of active debate for more than four decades~\cite{Plakhty2000, Yosephin1985, Quirion2006}. 

Geometric frustration also manifests in the physics of itinerant particles, for instance in the single-particle band structure: Similarly to spin systems, where the frustration is irrelevant for the ferromagnetic case but crucial for the antiferromagnetic one, the lowest band of the triangular lattice has a single minimum at the $\mathbf{\Gamma}$ point, but the frustration leads to two degenerate but distinct maxima at $\mathbf{K}$ and $\mathbf{K'}$ that correspond to the two opposite chiralities, see Fig.~\ref{fig:fig1}\textbf{c}.

In this work, we experimentally demonstrate the geometric frustration of the triangular lattice by realizing the superfluid-Mott insulator (SF-MI) transition at both the lower and upper band edges utilizing negative absolute temperature states in bosonic $^{39}$K to address the upper edge.  
In the frustrated case, weakly-interacting bosons are expected to condense into a chiral superfluid ($\chi$-SF) at nonzero momenta, namely the $\mathbf{K}$ or $\mathbf{K'}$ points at the corners of the Brillouin zone~\cite{Struck2011, OzawaPRR2023}. We experimentally demonstrate the lower stability of the $\chi$-SF  by observing a suppression of the critical value $|U/J|_\mathrm{c}$ for the transition to the MI by a factor of 2.7(3) and furthermore find different critical dynamics.

\section{Frustrated Bose-Hubbard model} \label{sec:GeomFrs}
Atoms in the lowest band of an optical lattice are described by the single-band Bose-Hubbard Hamiltonian~\cite{Jaksch_PRL_1998}: 
\begin{equation} \label{eq:BH_Ham_main} 
    \begin{split}
        \hat{H}_\mathrm{BH} = -J \sum _{\left\langle \mathbf{i,\,j} \right\rangle} {\hat b}_\mathbf{i}^\dagger{\hat b}_\mathbf{j}&+\frac{U}{2}\sum_\mathbf{i}{\hat n}_\mathbf{i}({\hat n}_\mathbf{i}-1) \\
        &- \sum_\mathbf{i}\left[\mu - V_\mathrm{ext}\left(\mathbf{i}\right)\right]\hat{n}_\mathbf{i}, 
    \end{split}
\end{equation} 
where ${\hat b}_\mathbf{i}$ and ${\hat b}_\mathbf{i}^\dagger$ are the annihilation and creation operators at site $\mathbf{i}$, ${\hat n}_\mathbf{i}= {\hat b}_\mathbf{i}^\dagger{\hat b}_\mathbf{i}$,  and ${\left\langle \mathbf{i,\,j} \right\rangle}$ denotes nearest-neighbour sites with tunnelling amplitude $J>0$. The on-site interaction energy is $U$, the chemical potential is denoted by $\mu$, and $V_\mathrm{ext}\left(\mathbf{i}\right)$ is the slowly-varying external potential due to the Gaussian nature of the lattice beams and the optical dipole trap.

\begin{figure*}[!htb] 
	\includegraphics[width=1\textwidth]{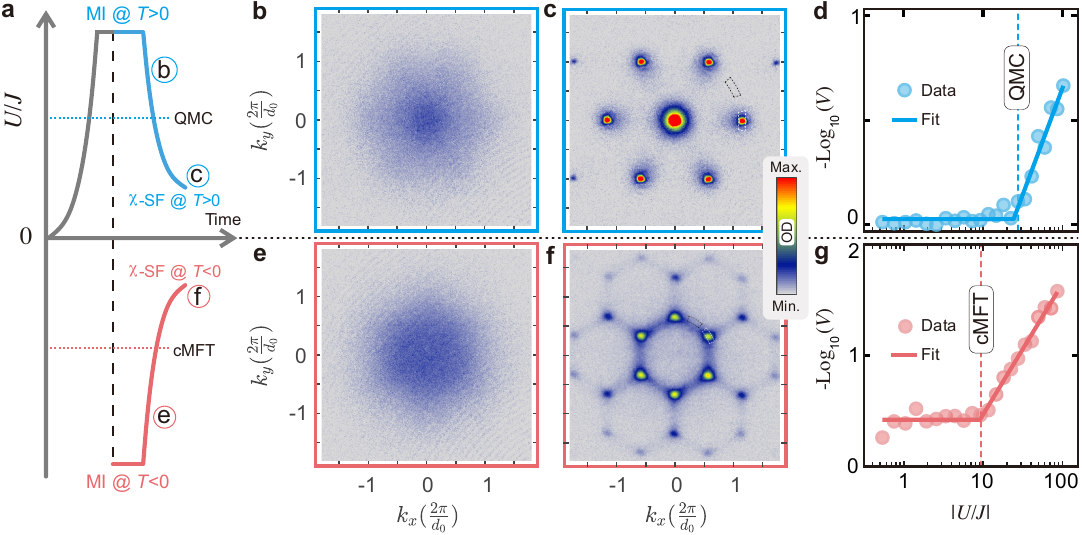}  
	\caption{ \textbf{($\chi$-)Superfluid-Mott insulator transition in \ufrs and \frs lattices}. \textbf{a},
 After creating a deep Mott insulator at $U>0,\,T>0$, we either continue with the positive temperature state (upper row), or flip the sign of $U$ (lower row) to create a negative temperature state. \textbf{b \& e}, The momentum distribution of atoms in a Mott insulator at lattice depth $22.5\,E_\mathrm{R}$ corresponding to $U/J\approx 200$ ($U/J\approx -200$) of the \ufrs (frustrated) system with $T>0$ ($T<0$). 
 \textbf{c \& f}, Momentum distribution of atoms in the superfluid ($\chi$-SF) regime at lattice depth $3\,E_\mathrm{R}$ corresponding to $U/J\approx 0.5$ ($U/J\approx -0.5$) of the \ufrs (frustrated) system, measured following a `booster' pulse, see Methods. 
 \textbf{d \& g}, Extracted visibility $V$ 
 of the first-order (zeroth-order) Bragg peaks for various final lattice depths at fixed scattering lengths $a_s = \pm 50\unt{a_0}$. The location of the kink in the fitted piecewise linear function shown by the solid blue (red) line yields the critical interaction $|U/J|_\mathrm{c}$ of the phase transition for the \ufrs (frustrated) system.  
 The dashed vertical lines in \textbf{d,g} indicate the transition points predicted by quantum Monte Carlo simulations (for the \ufrs system) and cluster mean-field calculations (for the \frs system) for homogeneous systems, see Methods.
 Each data point in \textbf{d} and \textbf{g} is an average of two repetitions of the same experiment, see Methods for details and raw data. 
 } \label{fig:fig2} 
\end{figure*} 
	
For weak interactions ($U/J \ll 1$),  the ground state is a superfluid, i.e., a Bose-Einstein condensate where the particles condense at the $\mathbf{\Gamma}$ point at the centre of the Brillouin zone, such that the wavefunction has long-range coherence with the same phase at every lattice site. For high densities and weak-but-finite interactions, the system can be mapped to the ferromagnetic case ($J_\mathrm{XY}>0$) of the classical $\mathrm{XY}$ model~\cite{Eckardt2010, Struck2011}, where the local phase at each lattice site is mapped onto a classical 2D spin. The above superfluid state directly corresponds to the ferromagnetic phase where all spins are pointing in the same direction, indicated by parallel blue arrows in the inset of Fig.~\ref{fig:fig1}\textbf{c}. 

In the antiferromagnetic case with $J_\mathrm{XY}<0$, on the other hand, the spins cannot anti-align on all bonds due to the geometric frustration and the classical $\mathrm{XY}$ model minimizes its energy by forming one of two chiral configurations where neighbouring classical spins are oriented at $120^\circ$ relative to each other, similarly to the Heisenberg model, see the inset of Fig.~\ref{fig:fig1}\textbf{c}. 
In the band structure, these two chiral phase patterns correspond directly to the two degenerate maxima at $\mathbf{K}$ and $\mathbf{K'}$, see Fig.~\ref{fig:fig1}\textbf{b}. They are time-reversal partners of each other and are not connected by a reciprocal lattice vector.  

As a consequence, condensation in the frustrated Bose-Hubbard model with $J<0$ can occur at not only one, but two distinct momenta, namely $\mathbf{K}$ and $\mathbf{K'}$. Any superposition of these two momenta would result in a stripe phase with a periodic density modulation, which, for weakly repulsive bosons, would increase the overall energy of the condensate. This corresponds to the immiscible regime in the context of spin-orbit coupled condensates~\cite{Wang2010,Ho2011}. Therefore, to reduce the interaction energy, the weakly-interacting system spontaneously condenses at either of the two points instead of creating a superposition~\cite{Becker2010, Struck2011, OzawaPRR2023}.  This results in a \textit{chiral superfluid} ($\chi$-SF) that breaks two distinct symmetries, namely the usual $\mathrm{U}(1)$ symmetry connected to condensation and the chiral $\mathbb{Z}_2$ symmetry~\cite{Zaletel2014}.   

For strong interactions, that is ${U/|J| \gg 1}$, the interaction dominates in both \frs and \ufrs cases and leads to bosonic Mott insulators (MI) at commensurate fillings~\cite{Thomas2017, Becker2010} without any broken symmetries. In analogy to the magnetic case, this opens two possibilities for the transition from a MI into a $\chi$-SF: both $\mathrm{U}(1)$ and $\mathbb{Z}_2$  symmetries might be broken simultaneously in a potentially discontinuous direct transition; alternatively, there could be an intervening chiral Mott insulator phase ($\chi$-MI) that breaks only the chiral $\mathbb{Z}_2$ symmetry and would be connected to the  $\chi$-SF via a continuous phase transition~\cite{LandauV5, Zaletel2014}. Although numerical works hint at the latter scenario, this question has not been settled~\cite{Zaletel2014, Lauchli2018}.

In optical lattices, the frustrated case is traditionally accessed either by lattice modulation (Floquet engineering) to flip the sign of the hopping $J$~\cite{Eckardt2010, Struck2011, OzawaPRR2023}, by leveraging artificial gauge fields in a synthetic dimension~\cite{Li2023}, or by loading atoms to higher orbital bands~\cite{SoltanPanahi2011, Jin2021}. These approaches are either accompanied by Floquet heating, constrained by the finite size of the lattice along the synthetic dimension, or limited by the finite lifetime of atoms in the higher bands. 
To move beyond these limitations, we instead access the frustrated case by realizing negative absolute temperatures that result in the stable occupation of the highest-energy state of the lowest band without any inherent heating~\cite{Purcell1951, Rapp2010, Braun2013}.  Intuitively, negative temperature states can be understood by considering the simultaneous flipping of the signs of the Hamiltonian $\hat{H}\to-\hat{H}$ and of the absolute temperature $T\to -T$, which will leave any thermal density matrix unchanged:
\[ \hat{\rho}=\exp\left(-\dfrac{\hat{H}}{k_B T}\right)=\exp\left(-\dfrac{-\hat{H}}{-k_B T}\right)\]
It directly follows that the complete phase diagram of $\hat{H}$ at positive temperatures $T\geq 0$ is identical to the phase diagram of $-\hat{H}$ at negative absolute temperatures $T\leq 0$. One well-known consequence of this general identity is that the highest excited state of a ferromagnetic spin model corresponds to the ground state of the corresponding antiferromagnetic model. Negative absolute temperatures are stable thermodynamic solutions that exist for any Hamiltonian whose energy density is locally bounded from above; in particular they exist for single-band lattice models with suitable interactions such as the attractive Bose-Hubbard model~\cite{Braun2013}.

\begin{figure*}[!ht] 
	\centering 
	\includegraphics[width=1\textwidth]{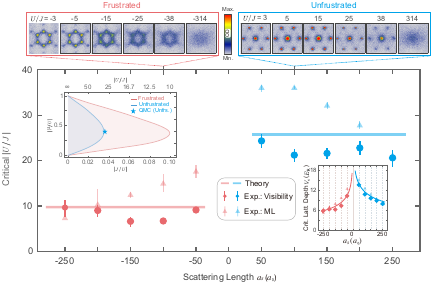} 
	\caption{ \textbf{ Frustration-induced suppression of superfluid order.} 
    Measured  critical interaction strengths $\left|\frac{U}{J}\right|_c$ for \frs (red) and \ufrs systems (blue) extracted using fits to the visibility as in Fig.~\ref{fig:fig2} (circles) and, independently, using a machine-learning (ML) approach (triangles). The average critical value extracted from the visibilities is $21.9\pm0.7$ for the \ufrs system, in stark contrast with the frustrated system with $8.2\pm0.5$, demonstrating the suppression of the $\chi$-SF  in the frustrated case compared to the SF in the \ufrs system. The blue line represents the Quantum Monte Carlo (QMC)  prediction of $\left(\frac{U}{J}\right)_\mathrm{c} = (25.5-26)$ for a homogeneous \ufrs system, while the red line represents the cluster mean-field prediction of critical $\left|\frac{U}{J}\right|_\mathrm{c} = 9.5$ for a \frs homogeneous system, see Methods. Right inset: Corresponding critical lattice depths at various scattering lengths. Left inset: Simulated phase diagram for homogeneous \frs and \ufrs  systems around the $n=1$ Mott lobe, calculated using cluster mean-field theory, see Methods. The prediction from QMC is shown as a blue star. Top row images: Measured momentum distributions (with `booster' pulse) for different $U/J$.
    } \label{fig:fig3} 
\end{figure*}

As a consequence, the phase diagram of the attractive Bose-Hubbard model at negative temperature ($U<0$, $T<0$) with the conventional hopping $J>0$ directly corresponds to the frustrated case ($J<0$) of the repulsive model at positive temperature. 
Thus, by realizing the attractive Hubbard model at negative temperature, we gain experimental access to the frustrated case that contains a chiral superfluid as the "ground state" without having to change the sign of $J$ and can hence avoid the associated heating.

\section{Experimental Setup} \label{sec:ExpSetup} 
Our experiment begins with the preparation of a Bose-Einstein condensate of approximately $90{-}110\times 10^3$ $^{39}\mathrm{K}$ atoms with no discernible thermal fraction in a crossed-beam dipole trap, see Methods for details. Subsequently, atoms are loaded into a three-dimensional optical lattice consisting of a one-dimensional, retro-reflected lattice of wavelength $\lambda_z = 1064\unt{nm}$ along the vertical $(z)$ direction, and a two-dimensional, triangular lattice in the horizontal $(xy)$ plane. To prevent tunnelling along the $z$ direction and create independent horizontal 2D layers, we use a large lattice depth $V_z = 26(1)\,E_\mathrm{R,1064}$ where $E_\mathrm{R,1064} = \frac{h^2}{2m\lambda_{z}^2}$ is the photon-recoil energy of a $^{39}\mathrm{K}$ atom with mass $m$  and $h$ is Planck's constant.  The horizontal triangular lattice consists of three interfering  beams of wavelength $\lambda_0 = 532\unt{nm}$ with in-plane polarization propagating at $120^\circ$ relative to each other in the $xy$-plane, as shown in Fig.~\ref{fig:fig1}\textbf{a}.  
We create a deep MI at unit filling ($n=1$) by loading the atoms at an $s$-wave scattering length of $a_s = 200~a_0$, where $a_0$ denotes Bohr's radius, into a triangular lattice of depth $V_0 = 25.5\,E_\mathrm{R}$. The recoil energy for the triangular lattice is defined as $E_\mathrm{R} = \frac{h^2}{2m\lambda_0^2}\frac{3}{4}$, where the factor $\frac{3}{4}=\sin^2\left({2\pi/3}\right)$ reflects the angle between the lattice beams, see Methods.

\section{Probing the ($\chi$-)SF--MI transition} \label{sec:SfMiProbe}
The above initial deep MI at unit filling with $|U/J|$ large compared to the critical value of the phase transition $|U/J|_\mathrm{c}$ is the key gateway enabling us to switch between positive and negative temperatures.
In the atomic limit ($J\rightarrow 0$), it can be approximated as a product of local Fock states and is at the same time the ground state of the repulsive Hubbard model and the highest excited state of the attractive one.
After preparing such a deep MI at positive temperature in the repulsive case, we take one of the two routes shown in Fig.~\ref{fig:fig2}\textbf{a}:  we either continue with the \ufrs positive temperature state (upper row) at repulsive interactions, or access the frustrated case by creating a negative temperature state (lower row) at attractive interactions~\cite{Braun2013}. For the latter case, we flip the sign of the interactions using the broad Feshbach resonance of $^{39}\mathrm{K}$ atoms at $402.74\,$G~\cite{Etrych2023} and change the sign of the underlying trapping potential $V_\mathrm{ext}(\mathbf{i})$ using an additional blue-detuned beam, see Methods. 

In both cases we then probe the transition from the MI into the (chiral) superfluid and extract the critical values of $|U/J|_\mathrm{c}$. To this end, we first ramp the magnetic field linearly to the required value in $4\unt{ms}$ to set the desired scattering length $a_s$. Afterwards, we decrease the depth of the triangular lattice to the desired value $V_0$ at a fixed rate of $0.43\,E_\mathrm{R}/\mathrm{ms}$, while keeping the vertical lattice at its original high depth. 
Finally, the coherence of the resulting state is probed by quickly increasing (`boosting') the depth of the triangular lattice within a few tens of $\mu$s to compress the Wannier functions without altering the quasi-momentum distribution~\cite{Stoferle2004,Yu2024}, before probing the momentum distributions using absorption imaging after $6\unt{ms}$ time of flight (TOF), see Methods.

For $|U/J|\gg 1$, the system remains Mott insulating with only small nearest-neighbour correlations, giving rise to the broad momentum distributions in Fig.~\ref{fig:fig2}\textbf{b,e}. The opposite signs of the nearest-neighbour correlations in the two cases are reflected in the subtly different patterns.
Once the system reaches the superfluid or $\chi$-SF regimes at weak interactions, in contrast, well-defined sharp Bragg peaks appear, see Fig.~\ref{fig:fig2}\textbf{c,f}, that demonstrate the emergence of longer-range coherence. Here the observed peak width is dominated by the in-situ cloud size in combination with the finite TOF~\cite{Braun2015}. 

The more prominent lines connecting the peaks at $\mathbf{K}$ and $\mathbf{K'}$ in the frustrated case are due to the smaller curvature of the band structure along the $\mathbf{K}-\mathbf{M}-\mathbf{K'}$ edge of the Brillouin zone, see Fig.~\ref{fig:fig1}\textbf{b}. This leads to the energy difference between $\mathbf{K,K'}$ and $\mathbf{M}$ being only 1/9 of the total bandwidth. 
We observe almost equal occupation of both $\mathbf{K}$ and $\mathbf{K'}$ even though the $\chi$-SF is expected to always spontaneously select one of the two possibilities.
This difference is partially due to the spontaneous symmetry breaking happening independently in each horizontal layer. Furthermore, the chosen ramp times are likely too fast for coherence to emerge over the full layer and instead lead to finite chiral domains, see below.

In order to locate the phase transition point, we extract the visibility of the first(zeroth)-order Bragg peaks by summing the density inside six annular arcs centred around the Bragg peaks --- white dashed boxes in Fig.~\ref{fig:fig2}\textbf{c} --- and comparing it to equal-area regions (black dashed boxes) that are $30^\circ$ away from the Bragg peaks, see Methods.
The resulting visibilities are shown by the blue and red data points in Figs.~\ref{fig:fig2}\textbf{d,g} and the critical values of $\left|U/J\right|$ for the two phase transitions can be extracted phenomenologically from the location of the kink in piecewise linear fits~\cite{Gerbier_PRL_2005, Becker2010}.

\begin{figure*}[!htb]
	\includegraphics[width=1\textwidth]{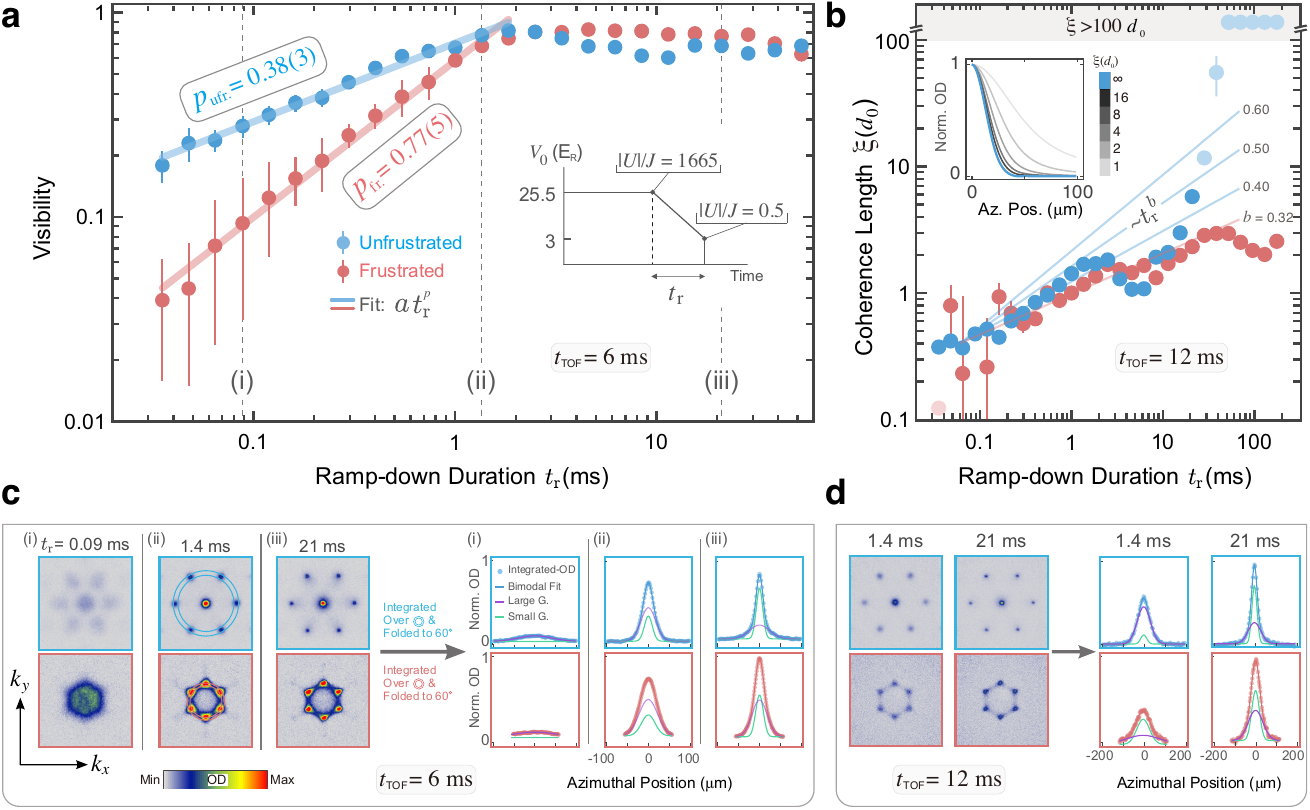}  
	\caption{\textbf{Emergence of coherence in \frs and \ufrs systems.} \textbf{a}, Starting from a deep Mott insulator with $\left|U/J\right|_\mathrm{i} \approx 1650$, we ramp into the weakly-interacting (superfluid) regime with $\left|U/J\right|_\mathrm{f} = 0.5$ by linearly reducing the lattice depth over a variable ramp-down duration $t_\mathrm{r}$, as shown in the inset, and measure the visibility at the constant final lattice depth. 
    The initial deep MIs have almost no coherence, resulting in low visibilities $V$. During the ramp-down, coherence smoothly builds up, with slower ramps leading to longer-range coherence, which causes a continuous increase of visibilities for both \frs and \ufrs cases. 
    The solid red (blue) lines show power-law fits  $at_\mathrm{r}^p$, where the exponents $p$  differ by a factor of $\approx2$ between the \frs and \ufrs case. 
    \textbf{b}, Coherence length $\xi$ extracted from time-of-flight pictures measured following an identical sequence to \textbf{a} but with a longer time of flight of 12 ms, see Methods. The inset shows azimuthal cuts through simulated time-of-flight images for various coherence lengths $\xi$ using the experimental in-situ size and time of flight. They highlight how for large $\xi\gtrsim10~d_0$, the peak shape becomes dominated by the in-situ density distribution~\cite{Braun2015}, leading to large systematic uncertainties in $\xi$. Hence, the corresponding data points (faded colours) are less reliable.
    \textbf{c}, (Left panel) Example time-of-flight images (6 ms TOF) for the data in \textbf{a} at three different ramp times $t_\mathrm{r}$ indicated by vertical dashed lines in \textbf{a}, for both \frs and \ufrs cases. (Right panel) Integrated density extracted by first integrating radially over a ring (shown schematically in (Left, ii)), and subsequently summing the six $60^\circ$ azimuthal segments, see Methods for details. The light-blue and light-red solid lines are bimodal fits, consisting of the sum of two Gaussians with different widths (green \& purple). \textbf{d}, Same as \textbf{c}, however at $t_\mathrm{TOF} = 12~\mathrm{ms}$ and the corresponding $t_\mathrm{r}$ are in the title of each figure. 
    } \label{fig:fig4} 
\end{figure*}

Figure~\ref{fig:fig3} presents the extracted critical values $|U/J|_c$ for various scattering lengths in both \frs (red) and \ufrs (blue) cases, revealing a stark contrast that underscores the impact of frustration: the superfluid in the \frs system is markedly more fragile compared to its \ufrs counterpart and the critical interaction strength  $|U/J|_\mathrm{c}$ is reduced by a factor of $2.7(3)$. This increased sensitivity to interactions is directly analogous to the suppression of antiferromagnetic order in frustrated magnets. It is directly visible already in single time-of-flight images (Fig.~\ref{fig:fig3}, top insets), where the Bragg peaks in the \frs case broaden and disappear at far lower interactions  than in the \ufrs case.

The measured critical values are compared with Quantum Monte Carlo simulations (QMC)~\cite{Pollet2012} for the \ufrs case, and cluster mean-field calculations (cMFT)~\cite{Schmidt2021} for the \frs case where QMC cannot be used, see Methods. The cMFT phase boundary of the $n = 1$ Mott lobe is shown in the left inset of Fig.~\ref{fig:fig3}  along with the QMC prediction (blue star) for the \ufrs case.

To ascertain the systematic uncertainties inherent in the above phenomenological approach of locating the kink of the visibility, we compare the results with a supervised machine-learning approach, where the strong first-order peaks have been masked to force the network to focus on information complementary to the prominence of the peaks, see Methods.  
Even though the two methods result in slightly different critical values, they both independently confirm the suppression of superfluidity in the \frs case.

\section{Emergence of coherence in a \frs triangular lattice}
 
We now turn to the dynamics of the $\chi$-SF to MI transition in the \frs system and compare it to the known continuous transition of the \ufrs system~\cite{Fisher1989}. In order to observe how coherence emerges, we start from deep Mott insulators in both \frs and \ufrs lattices prepared similarly to the above. We then cross the transition and ramp to a fixed lower final value of $\left|U/J\right|_f=0.5$ in the superfluid ($\chi$-SF) regime by linearly decreasing the lattice depth in a variable ramp-down time $t_{\mathrm{r}}$, see inset in Fig.~\ref{fig:fig4}\textbf{a}. We extract both the visibility (Fig.~\ref{fig:fig4}\textbf{a}), which can capture the dynamics for short ramp times, as well as the averaged coherence length $\xi$ (Fig.~\ref{fig:fig4}\textbf{b})~\cite{Braun2015} that is extracted using fits to simulated data, see Methods for details. 

Both visibility and averaged coherence length show that the coherence of the final system increases continuously with increasing ramp time. This is in contrast to the quantum metastability expected at a \textit{strongly} discontinuous transition characterized by abrupt changes in the order parameter~\cite{Trenkwalder2016, Song2022}.

At the same time, heuristic power-law fits to the visibility result in significantly different exponents, hinting towards different critical dynamics.  
Even more strikingly, for the slowest ramps, the \ufrs case results in very sharp peaks that are consistent with the emergence of long-range order spanning the whole system, while the coherence length in the frustrated case never grows above $\approx 3\, d_0$. This is qualitatively consistent with very recent approximate simulations using the discrete truncated Wigner approximation~\cite{nagao2025}. The oscillations visible in the coherence length are probably related to breathing dynamics in the presence of the harmonic trap~\cite{Braun2015}.
Simple Kibble-Zurek scaling arguments predict that, for very slow ramps, the coherence length in the \ufrs case should grow as a power law with an exponent of $b_\text{ufr.}=0.4$  governed by the critical exponents of the 3DXY transition, see Methods. However, for short ramp times, previous measurements in a square lattice found exponents around $\tilde{b}_\text{ufr.}\approx0.6{-}0.8$~\cite{Braun2015}. Barring the oscillations due to breathing dynamics, and taking into account that the measured coherence lengths become less reliable for very large coherence lengths where the TOF profile becomes dominated by the in-situ profile, see inset, the observed dynamics becomes consistent with a similar power law.
The example pictures in Fig.~\ref{fig:fig4}\textbf{c}\&\textbf{d} show the emergence of a strongly bimodal distribution in the \ufrs case,  highlighting the emergence of long-range coherence. Similarly to the growing but limited coherence length in the \frs case, the corresponding pictures show less pronounced signs of bimodality.

The observed dynamics leave several possibilities regarding the nature of the \frs phase transition: It could be weakly first order, where the observed emergence of short-range coherence occurs within the MI while approaching the phase transition. Or it could be continuous but with rather different critical exponents compared to the \ufrs case. An especially intriguing possibility is a sequential transition with an intervening chiral Mott insulator ($\chi$-MI)~\cite{Zaletel2014, Lauchli2018}. In this scenario, the transition from $\chi$-MI to $\chi$-SF would be of the same 3DXY type as the transition in the \ufrs case, but the achievable coherence length would be limited by the domain size of the preceding chirality-breaking transition from MI to $\chi$-MI and the growth of the coherence length would be governed by the critical exponents $b_I\approx 0.32$ of the corresponding Ising universality class (see Methods), indicated by the red line.

\section{Conclusion}
In conclusion, we have experimentally studied the superfluid-Mott insulator transition in a triangular optical lattice for both \ufrs and \frs cases. By preparing long-lived negative-temperature states, we gain access to the frustrated band maxima at $\mathbf{K}$ and $\mathbf{K'}$ without Floquet driving or higher-band loading. Relative to condensation at $\mathbf{\Gamma}$, the superfluid regime is partially suppressed in the \frs case, and the critical interaction for the transition into the Mott insulator $|U/J|_c$ is reduced by a factor $2.7(3)$. The unfrustrated phase boundary agrees with quantum Monte Carlo benchmarks, while the frustrated boundary is consistent with cluster mean-field calculations. The substantial suppression underscores the role of lattice geometry and frustration in quantum phase transitions and is directly reminiscent of the suppression of antiferromagnetic order in frustrated magnets. 

We furthermore probe the nature of the phase transition,  which in the \frs system involves breaking of the chiral symmetry in addition to the usual U(1) symmetry. Although both \frs and \ufrs transitions show a continuous growth of coherence for fast ramps, we find a significant contrast in the emergence of coherence for slower ramps. While we observe system-size coherence in the \ufrs case, in the frustrated case the coherence length saturates to only a few lattice constants for the used ramp-times, highlighting the different critical behaviour of this more complex phase transition. 

The application of negative temperature states for accessing the frustrated regime of a lattice system opens new avenues for quantum simulations of frustrated systems, such as the kagome lattice, where traditional approaches often fail. The critical understanding of how frustration influences interactions and phase stability could facilitate the design of quantum materials and devices, and the methods employed here can be adapted to explore other complex systems where geometric and interaction-induced frustrations play a critical role.

\clearpage
\section{Methods} 
\subsection{Bose-Einstein condensate} 
The preparation of the \({}^{39}\)K Bose-Einstein condensate (BEC) starts by loading laser-cooled \({}^{87}\)Rb and \({}^{39}\)K atoms into a quadrupole magnetic trap followed by forced microwave evaporation to $7\text{-}10\,\mu$K. Using a focus-tunable lens (Optotune EL-16-40-TC) to translate the focus of a $7\,$W beam at 1064 nm with a waist of 50 $\mu$m, both species are then optically transported~\cite{Leonard_2014} in $2.9\,$s over $50\,$cm into the science chamber. 
We typically load up to $15\times 10^6$  \({}^{39}\)K and up to $20\times 10^6$ \({}^{87}\)Rb into the transport beam and transport about 60-85\% of the atoms while the temperature increases by about 35\%. 

Once in the science chamber, the atoms are loaded into a crossed-beam optical dipole trap (XODT) formed by two horizontal beams of $1064\,$nm light with beam waists of $240\times55~\mu\mathrm{m}^2$. Evaporation continues by lowering the depth of the dipole trap and we ensure good thermalization between the two species by setting the interspecies scattering length to $140\,a_0$  using the ${}^{87}$Rb-${}^{39}$K Feshbach resonance at 317.9 G~\cite{Simoni2008}.

Due to gravity and the mass ratio of $m_{Rb}/m_{K}\approx2.2$, rubidium atoms are preferentially lost during the evaporation and thereby act as a coolant for the potassium. After losing all ${}^{87}$Rb, we continue evaporation with ${}^{39}$K alone at an intra-species scattering length of $150\,a_0$. 
This leads to an almost pure BEC of $90-110\times 10^3$ ${}^{39}$K atoms in the $\ket{F,m_F} = \ket{1, 1}$ state with no discernible thermal fraction. The interaction between potassium atoms is controlled using the broad Feshbach resonance at 402.74 G~\cite{Etrych2023} by applying a homogeneous magnetic offset field $B_z$ along the vertical direction.

\subsection{Triangular lattice} 
The triangular optical lattice in the horizontal plane is created by overlapping three single-frequency laser beams with a common wavelength of $\lambda_0 = 532\, \mathrm{nm}$. The beams propagate at $120^\circ$ relative to each other with in-plane polarization and intersect at the location of the atoms, see Fig.~\ref{fig:fig1}\textbf{a} of the main text. 

Ground state atoms are repelled by the blue-detuned $532\,$nm light and reside at the intensity minima of the triangular optical lattice. The resulting potential is given by 
\begin{align}\label{eq:FullLatticePotential}
    \begin{split}
        V({\textbf r}) =V_0\Big( \frac{2}{3} - \frac{2}{9} \Big(&\cos[{\textbf G_1}\cdot{\textbf r}] + \cos[{\textbf G_2}\cdot{\textbf r}]    \\  
        & +\cos[({\textbf G_1}+{\textbf G_2})\cdot{\textbf r}] \Big)\Big),
    \end{split}
\end{align}
where ${\textbf G_{1,2}} = \left(\frac{1}{\sqrt{3}},\,\pm1\right)\frac{2\pi}{d_0}$ are the reciprocal lattice vectors, and $d_0 = \frac{2}{3}\lambda_0$ denotes the lattice constant, as shown in Fig.~\ref{fig:fig1}\textbf{a} of the main text. The lattice depth $V_0$ is defined as the total range of the potential. The real-space lattice vectors, denoted as ${\textbf a}_{1,2}$, are derived from the lattice potential in Eq.~\ref{eq:FullLatticePotential} and are  given by ${\textbf a}_{1,2} = \left(\frac{\sqrt{3}}{2},\pm \frac{1}{2}\right)d_0$. 

The orthogonal vertical optical lattice is a red-detuned retro-reflected 1D lattice formed by light of wavelength $\lambda_z = 1064\,\mathrm{nm}$. The potential along the vertical $z$-direction reads 
\begin{align}\label{eq:VzPotential}
    \begin{split}
        V(z) = V_z\cos^2\left(k_z z\right),
    \end{split}
\end{align}
where $V_z$ is the total range of the vertical lattice potential, and $k_z = \frac{2\pi}{\lambda_z}$. 

Throughout the paper, we have used $E_\mathrm{R} = \frac{\hbar^2}{2m}\left|\frac{{\mathbf G}}{2}\right|^2 = \frac{3}{4}E_{\mathrm{R},\, 532\, \mathrm{nm}}$ as the unit of the lattice depth for the triangular lattice. Here, $E_{\mathrm{R},\, 532\, \mathrm{nm}} = \frac{\hbar^2k_0^2}{2m}$ represents the photon-recoil energy of an atom with mass $m$, and $k_0=2\pi/\lambda_0$. The unit of the vertical lattice depth is the corresponding photon-recoil energy $E_\mathrm{R,1064} = \frac{\hbar^2k_z^2}{2m}$.

From the lattice potentials in Eq.~\ref{eq:FullLatticePotential}~\&~\ref{eq:VzPotential} we calculate the band structure, the maximally localized Wannier functions, the nearest-neighbour hopping energy $J$, and on-site interaction energy $U$~\cite{Walter_PRA_2013}, taking into account the presence of the vertical lattice.

\subsection{Calibration of magnetic field and lattice depth} 

We calibrate the magnetic field $B_z$ using radiofrequency transfer of ${}^{87}$Rb atoms from the $\ket{F,m_F}=\ket{1,-1}$ state to $\ket{1,0}$ for a set of fixed field values and compare the extracted resonance frequencies to the Breit-Rabi formula. This provides a calibration of $B_z$ from a few Gauss up to beyond the Feshbach resonance at \SI{402.74}{G} with an overall uncertainty of \SI{60}{mG}. 

After aligning the three $532\,$nm lattice beams independently onto the atoms, we create three one-dimensional (1D) lattices from each pair of $532\,$nm beams. The lattice depths of each of the three 1D lattices are calibrated separately for several lattice depths using Kapitza-Dirac diffraction~\cite{Gadway2009}. This calibration yields an overall lattice depth uncertainty of 1-2\%.

\subsection{Experimental Sequence}
\begin{figure*}[t]
    \centering
    \includegraphics[width=0.9\textwidth]{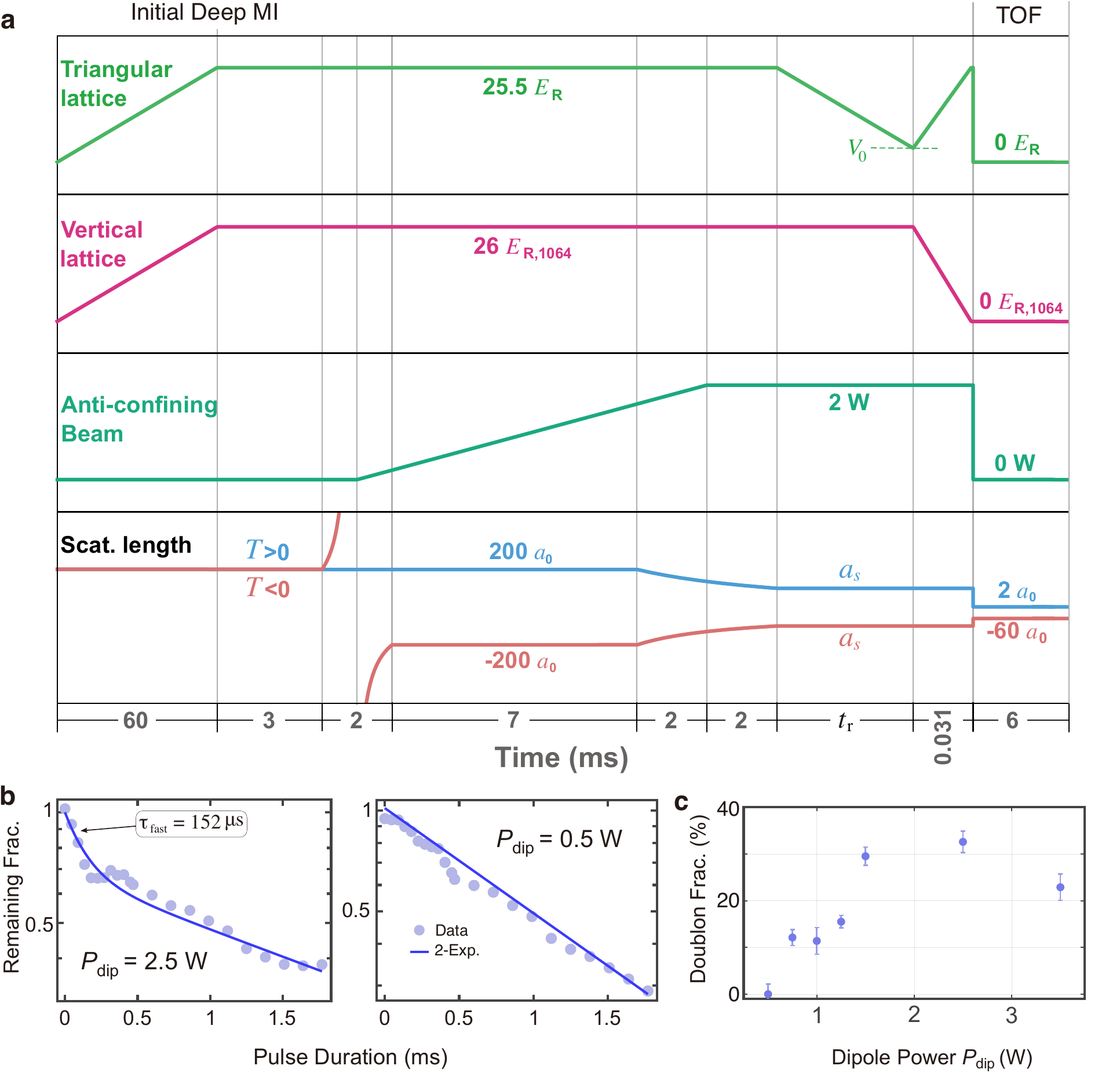}  
    \caption{\textbf{Experimental sequence; Doublon fraction in the initial deep Mott insulator.} \textbf{a}, The positive and negative temperature sequences share the same ramps of the vertical and triangular lattices. There is a common initial loading to a deep lattice. The anti-confining beam is used for negative temperature sequences only. For negative temperature sequences, the scattering length is ramped from positive to negative values (shown in red) while for positive temperature sequences the scattering length remains positive throughout (blue). \textbf{b}, Remaining atom number as function of the duration of a near-resonant light pulse together with two exponential fits. Higher occupancies ($n>1$) lead to a fast pair-wise loss on timescale $\tau_{\mathrm{fast}}$ and the amplitude of the fast decay measures the fraction of higher occupancies. \textbf{c}, The extracted doublon fraction as a function of the power of each XODT beam. The errorbars are one standard deviation of the fit uncertainty.   } \label{fig:seq_doublon} 
\end{figure*} 

The experimental sequence for the measurements presented in this paper is shown in Fig.~\ref{fig:seq_doublon}\textbf{a}. After the preparation of the BEC, we prepare a deep Mott insulator (MI) at a scattering length of 200 $a_0$ in the three-dimensional lattice by linearly ramping the vertical and horizontal lattice beams to a lattice depth of $V_z=26(1)\,E_\mathrm{R,1064}$ and $V_0=25.5\,E_\mathrm{R}$, respectively. 

For the \frs case, we first hold the atoms for \SI{3}{ms} in the deep MI and then create a negative temperature state by ramping the scattering length to $-200\,a_0$ in \SI{2}{ms}. Halfway through this ramp we start to introduce a blue-detuned 532-nm anti-confining beam along the vertical direction to flip the sign of the harmonic trap $V_{\mathrm{ext}}$ and thereby stabilize the negative temperature state~\cite{Braun2013}. For the \ufrs case, on the other hand, we remain in a positive temperature state where we neither ramp the scattering length nor introduce the anti-confining beam. 
In both cases, we then ramp the scattering length in \SI{4}{ms} to the desired value, $a_s$. 

We then finally probe the phase diagram by ramping down the horizontal lattice to the final lattice depth $V_0$ while maintaining a deep vertical lattice. This ensures that the system consists of a series of two-dimensional triangular lattices in the horizontal plane. For the data presented in Fig.~\ref{fig:fig4} the rate of this ramp is varied and for all other data sets the rate is set to match that of the initial ramp-up. 

For the measurement of the phase diagram (Figs.~\ref{fig:fig2} \& \ref{fig:fig3}), we finally perform a short `boost' ramp \cite{Stoferle2004, Yu2024} that increases the population of atoms in the satellite peaks for weak lattices and ensures that the Wannier state at the end of the sequence is the same for all lattice depths.  The `boost' ramp consists of ramping the horizontal lattice to $25.5\,E_\mathrm{R}$ in \SI{30}{\mu\second} and then holding for \SI{1}{\mu\second} followed by an instantaneous  switch-off of all lattice beams, anti-confining beam, and XODT. The chosen timescale is short enough such that the quasi-momentum distribution cannot change during the ramp.

Atoms are imaged using absorption imaging, after a time of flight at a scattering length of $2\,a_0$ (for $T>0$) or $-60\,a_0$ (for $T<0$). For most of the data, the TOF was $6\,$ms (Fig.~\ref{fig:fig2}, Fig.~\ref{fig:fig3}, and Fig.~\ref{fig:fig4} \textbf{a},\textbf{b}), except for the data in Fig.\ref{fig:fig4} \textbf{c},\textbf{d}, where the TOF was $12\,$ms. 

\subsection{Doublon fraction measurement}
 In order to exclude higher occupancies $(n>1)$, which would lead to inter-band excitations upon crossing the Feshbach resonance~\cite{Busch1998}, we measure the fraction of higher occupancies in the initial deep Mott insulating state using short pulses of near-resonant light with a detuning of $-40\,$MHz from the $D_2$ line, see Fig.~\ref{fig:seq_doublon}\textbf{b}. For sites with higher occupancies (left panel, strong XODT), pairs of atoms are quickly lost due to light-assisted collisions~\cite{DePue1999, Ronzheimer2013}. The atom number hence decays as the sum of two exponentials, and the number of atoms of higher occupied sites can be extracted from the amplitude of the fast decay~\cite{Ronzheimer2013}.

By changing the power of the XODT that operates concurrently with the lattice beams, we adjust the overall confinement and thereby control the central chemical potential of the system. With higher power, the trap becomes more confining, leading to the emergence of regions with higher occupancies within the Mott insulator, resulting in the characteristic ``wedding cake'' structure~\cite{Jaksch_PRL_1998}.  
The measured doublon fraction is depicted in Fig.~\ref{fig:seq_doublon}\textbf{c} as a function of dipole power.
For lower dipole powers up to  $P_\mathrm{dip} \approx 0.5\,\mathrm{W}$, the doublon fraction is insignificant and the decay profile is well fitted with a single exponential, as shown in the middle panel of Fig.~\ref{fig:seq_doublon}\textbf{b}.  For all the data presented in this work, the dipole trap power for each beam was set to 0.50 W to ensure that the initial MI has $n=1$.  

\subsection{Data analysis}
\subsubsection{Visibility Extraction}
\begin{figure*}[!htb] 
    \centering
    \includegraphics[width=1\textwidth]{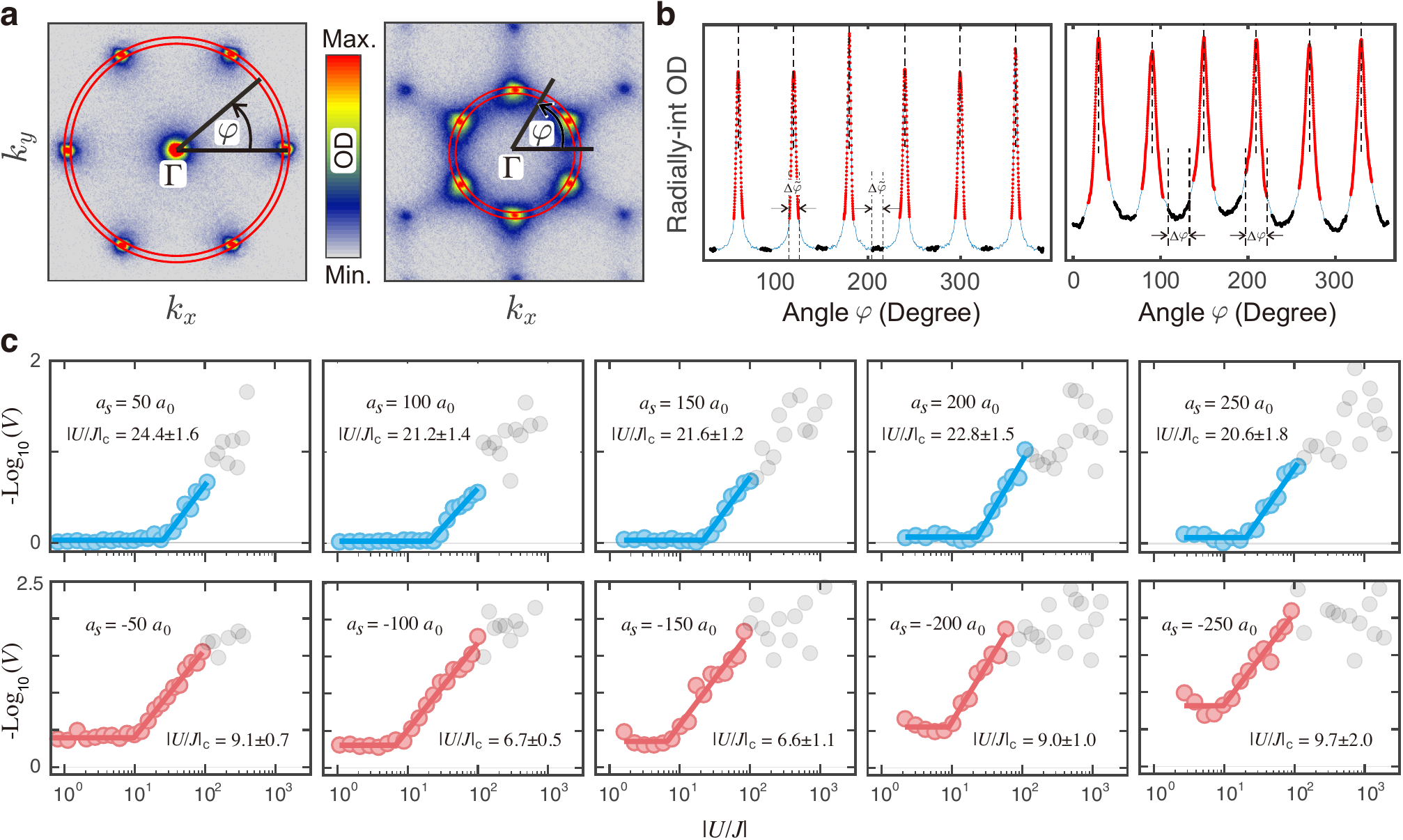}  
    \caption{\textbf{ Extraction of visibility and phase boundary.} \textbf{a}, Optical density (OD) of atoms in time-of-flight images for $T>0$ and $T<0$. The ring-shaped regions of interest are indicated by red circles. \textbf{b}, The corresponding angular distributions $\mathrm{OD}(\varphi)$  after performing a radial integration. The angular location of the peaks are independently extracted and remain fixed for a particular dataset.  
    The red (black) data points correspond to the regions  around the maxima (minima) that we sum over for extracting the visibility. 
    We use $\Delta {\Tilde{\varphi}}= 12^\circ$ and $\Delta\varphi=24^\circ$ and have checked that the results do not depend significantly on the chosen values. 
    \textbf{c}, Top row: Extracted visibilities across the SF-MI transition for the \ufrs system at positive scattering lengths $a_s$. The piecewise linear fits used to extract the critical values of $\left(\frac{U}{J}\right)$ are shown by the blue lines. We identify the phase transition with the location of the kink in each of the fits. Bottom row: Same analysis for the  \frs system. For both rows, the text inside each panel reports the extracted values of the critical $\frac{U}{J}$ along with the corresponding scattering length $a_s$. The grey data points are excluded from the fits. The reported uncertainties of the critical $\left(\frac{U}{J}\right)$ are one standard deviation of fit-uncertainty. }  \label{fig:vsblty_and_detail_of_PhaseDiag} 
\end{figure*} 

\label{subsec:visibiliDetail}
To extract the visibility~\cite{Gerbier_PRL_2005} from the optical density (OD) of time-of-flight images, we first identify the location of the satellite peaks and select a ring-shaped region of interest centred around the $\mathbf{\Gamma}$ point with different radii for $T>0$ and $T<0$,  as shown in Figs.~\ref{fig:vsblty_and_detail_of_PhaseDiag}\textbf{a}. Next, we radially integrate the OD within these rings to derive the angular distribution $\mathrm{OD}(\varphi)$, see Fig.~\ref{fig:vsblty_and_detail_of_PhaseDiag}\textbf{b}.  
We then integrate $\mathrm{OD}(\varphi)$ over intervals of width $\Delta\varphi$ around the maxima (red) and the halfway points between maxima (black),  yielding $\mathrm{OD}_\mathrm{R}$ and $\mathrm{OD}_\mathrm{B}$. The extracted visibility is defined as:

\begin{equation}
    V = \frac{ \mathrm{OD}_\mathrm{R} - \mathrm{OD}_\mathrm{B} }{ \mathrm{OD}_\mathrm{R} + \mathrm{OD}_\mathrm{B} }. \label{eq:Visibility}
\end{equation}

Note that the above definition of visibility directly yields the contrast of the structure factor $\mathcal{S}\left({\mathbf k}\right) = \sum_{\mathbf {i,j}} \left\langle {\hat b}_{\mathbf i}^\dagger {\hat b}_{\mathbf j}  \right\rangle\cdot$ $e^{i{\mathbf k}\cdot( {\mathbf r}_\mathbf{i} -{\mathbf r}_\mathbf{j} )}$, since the Wannier function of the triangular lattice has $\mathrm{C}_6$ symmetry and eventually becomes almost rotationally symmetric for large lattice depths.

Fig.~\ref{fig:vsblty_and_detail_of_PhaseDiag}\textbf{c} shows the extracted visibility, as a function of $U/J$, at various scattering lengths. Each panel corresponds to a data point in Fig.~\ref{fig:fig3} in the main text.  To extract the phase transition point $\left(U/J\right)_c$ from each data set, we plot the inverse of the visibility  on a log-log scale, fit it with a piecewise linear function, and heuristically identify the phase transition with the kink~\cite{Spielman2008, Becker2010}. 

\subsubsection{Coherence length extraction} \label{method:ChorenceExtraction} 

\begin{figure*}[!htb] 
    \centering
    \includegraphics[width=1\textwidth]{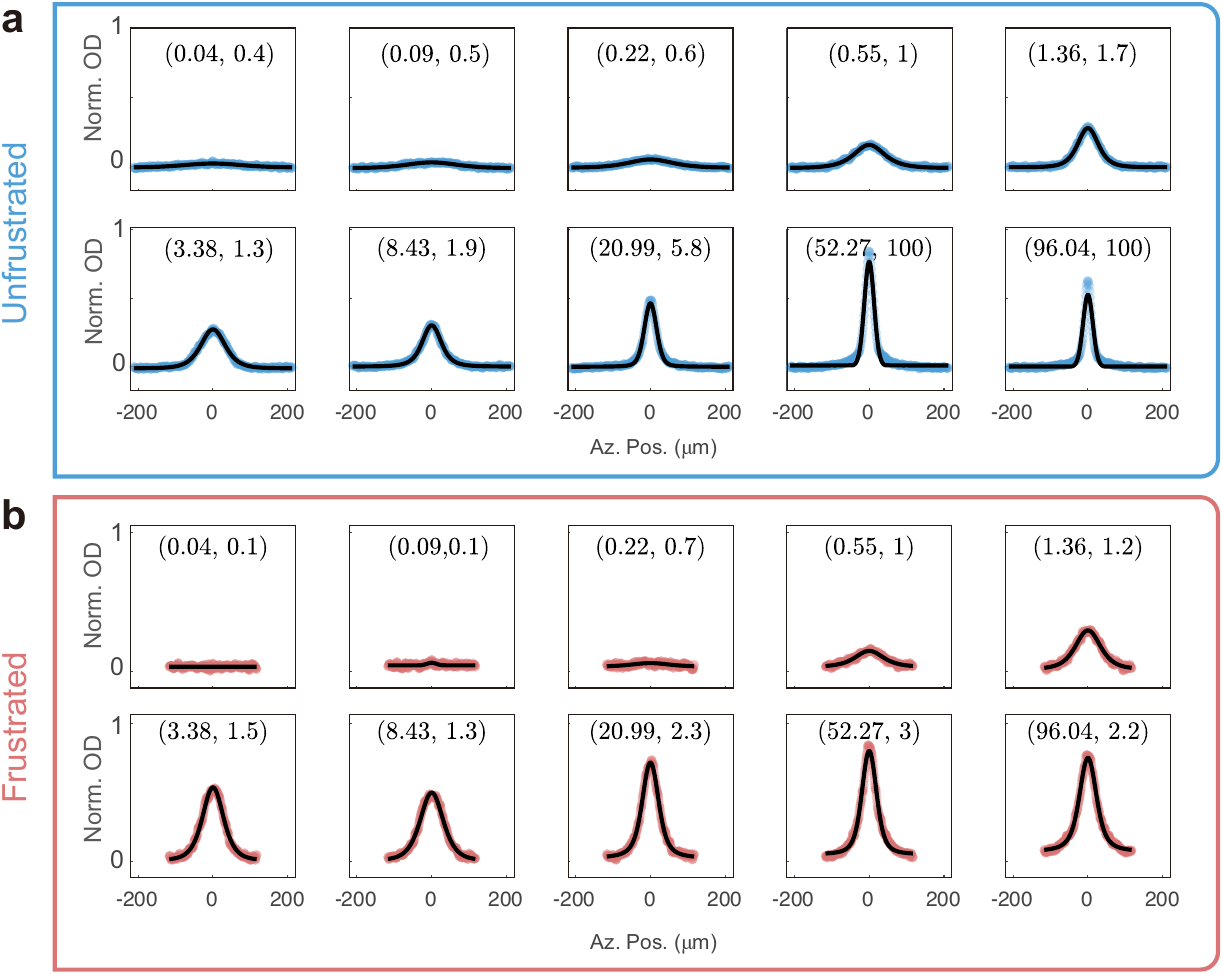}  
    \caption{\textbf{ Extraction of coherence length $\xi$ shown in Fig.~\ref{fig:fig4}\textbf{b}.} \textbf{a},  \ufrs case, the data in blue dots and the associated fit (line) to extract the coherence length $\xi$. The texts in each panel report ramp time and coherence length as $\left(t_\mathrm{r}\,\mathrm{(ms)},\,\xi\,(d_0)\right)$. \textbf{b}, same as \textbf{a}  for the \frs case. 
    }  \label{fig:coherence_len_extraction} 
\end{figure*} 
To extract the coherence length reported in Fig.~\ref{fig:fig4}\textbf{b} from the time-of-flight images, we follow the method used in~\cite{Braun2015}. Starting from images taken with a longer time of flight of $12\,$ms, we first extract the radially-integrated OD($\varphi$), similarly to the previous section.  Afterwards, we split the full $360^\circ$ curve into six equal segments and sum them up, see Fig.~\ref{fig:coherence_len_extraction}.

We then fit the resulting peak shapes to a list of simulated distributions assuming coherence lengths $\xi$ ranging from $0.1\,d_0$ to $100\,d_0$ and thereby extract the best-fitting coherence length $\xi$~\cite{Braun2015}. Due to the finite time of flight, the simulated peak shapes crucially depend on the in-situ size of the cloud, which was measured independently for each of the ramp-down durations $t_r$, for both \frs and \ufrs cases.

For large values of the coherence length $\xi\gtrsim 10\,d_0$, the measured distribution converges on the in-situ distribution~\cite{Braun2015}. This is demonstrated in the inset of Fig.~\ref{fig:fig4}\textbf{b} and leads to increasing uncertainties for large coherence lengths. In Fig.~\ref{fig:coherence_len_extraction}, we show the best-fitting distributions (solid black lines) together with the experimental data in blue (red) points for the \ufrs (frustrated) case shown in the top (bottom) panel. 

\subsubsection{Machine Learning}  
In addition to the established heuristic approach discussed above,
 we have also utilized a \textit{deep-learning} approach to estimate the phase boundary. The principle is as follows: Firstly, a deep neural network is trained with samples taken far away from the hypothesized phase boundary to distinguish both phases. Then, previously unseen samples taken at the intermediate parameter regime close to the phase transition are fed into the trained network. For these samples, the network then predicts $p_\text{MI}$, the probability for the sample to be in the Mott insulating regime. We then define the critical lattice depth at the point where $p_\text{MI}$ crosses $1/2$.  In order to extract complementary information to the visibility, we have masked the strong first-order peaks to force the network to focus on information complementary to the prominence of the peaks. 

To effectively handle the phase classification at both positive and negative temperatures, which exhibit distinct reciprocal space structures, we developed separate models for the two temperature regimes. For positive temperatures, we use a convolutional neural network (CNN) with three convolutional layers and two fully connected layers, totalling 6887 parameters. For negative temperatures, on the other hand, the network comprises three convolutional layers and one fully connected layer with 8188 parameters. Each network employs ReLU activation functions after each layer and a softmax output function to generate a probability distribution for the phases. Regularization is achieved through a dropout layer with an empirically determined rate of $p_{\mathrm{drop}}=0.4$ after each convolutional layer. These models are built using PyTorch and trained with the ADAM optimizer at a learning rate of $\gamma = 1 \times 10^{-3}$, $\lambda = 1 \times 10^{-4}$, batch size of 12, and concluding after 50 epochs. 

The robustness of our approach against training set limitations and biases was checked by training several models on various datasets, differing in size and minimum distance $\Delta$ from the phase boundary. These models consistently predict similar results, signalling the insensitivity of the outcome to specific training set characteristics.

\subsection{Stability of the Negative Temperature State}
\begin{figure*}[t] 
    \centering
    \includegraphics[width=1\textwidth]{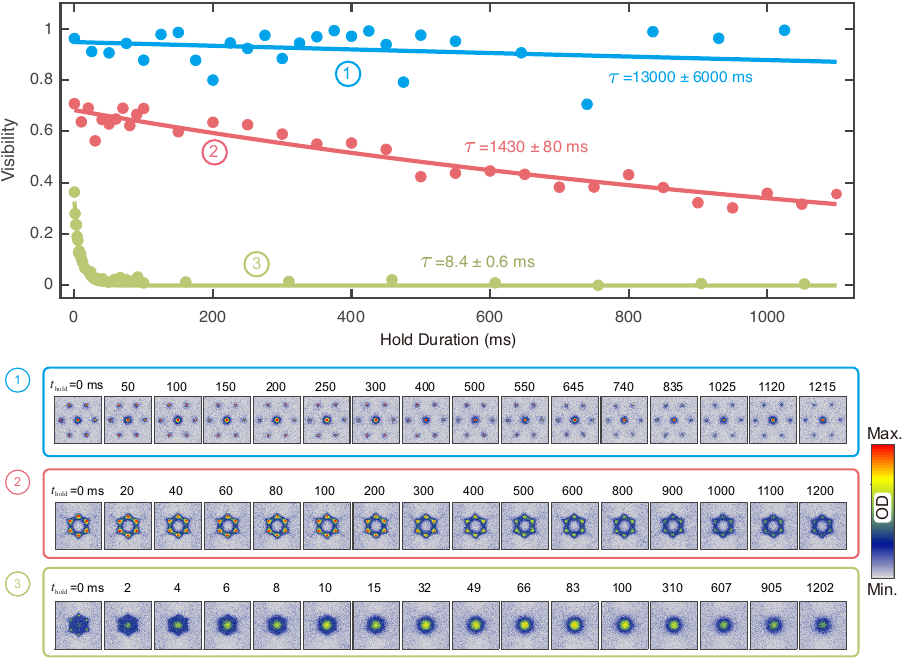}  
    \caption{\textbf{Stability of positive and negative temperature superfluids.} Shown are the three different situations where we probe the visibility of the Bragg peaks, along with the estimated value of the lifetime $\tau$ for an exponential fit $V(t) = V_\mathrm{max} \exp\left(-t/\tau\right)$, shown by the solid lines. The blue data points along with the corresponding fit represent the positive temperature case where atoms are held at $V_0 = 4.5 \unt{E_R}$ and $a_s = 60\unt{a_0}$. The red data points are for the corresponding negative temperature state with $a_s = -60\unt{a_0}$ and $V_0 = 4.5 \unt{E_R}$. The green data points represent the same negative temperature experimental condition as the red data except that we omit the anti-confining (AC) beam. The three rows of images are raw time-of-flight images for some of the data points shown in the main figure. The colours of the outline of each row correspond to the colour of the data points. } \label{fig:stability_neg_T} 
\end{figure*}  
We probe the stability of the negative temperature state by holding atoms in the frustrated superfluid regime for a variable time and observe the evolution of the visibility as a function of the hold duration, as shown by red data points in Fig.~\ref{fig:stability_neg_T}. The visibility $V$ is fitted with an exponential function $V \sim \exp(-t/\tau)$ to extract the lifetime $\tau$ of the visibility. The  resulting lifetime $\tau = 1430(80)\, \mathrm{ms}$ (at $a=-60\,a_0$) is about ten times shorter than in the corresponding positive temperature superfluid $\tau = 13(6)\, \mathrm{s}$ (at $a=60\,a_0$). 
However, the observed lifetime is still substantial and corresponds to about $4500$ tunnelling times, highlighting the stability of the negative temperature state.

As shown in Fig.~\ref{fig:fig1}\textbf{b} of the main text, the dispersion relation is rather flat along the edge of the Brillouin zone, such that the saddle points at momentum $\mathbf{M}$ have an energy of $8\,J$ and are hence significantly closer to the $\mathbf{K}$ and $\mathbf{K'}$ points (energy $9\,J$) than to the $\mathbf{\Gamma}$ point (energy $0\,J$). Assuming a slow heating process, this dispersion relation not only explains why the negative temperature distribution evolves towards a hexagonal ring, see Fig.~\ref{fig:stability_neg_T} (red), instead of directly filling the full Brillouin zone; it furthermore means that for the same effective heating rate, i.e., the same $|\dot{T}|$, the visibility of the frustrated (i.e.~$T<0$) states would also be expected to decay faster. 

Additionally, we examined the evolution of the visibility at negative temperature without the anti-confining beam, see the green data points and the corresponding fit in Fig.\ref{fig:stability_neg_T}. The lifetime in this scenario ($\tau = 8.4(6)\, \mathrm{ms}$) is nearly 200 times shorter than when using the anti-confining beam, highlighting the importance of flipping the sign of the external potential $V_{\mathrm{ext}}$ for $T<0$ states~\cite{Braun2013}.

\subsection{Calculation of the SF-MI phase transition point} \label{subsec:QMCcMFTDiscussion}
\subsubsection{Quantum Monte Carlo}
We calculate the superfluid-Mott insulator phase transition point at various temperatures  using the `worm' algorithm for quantum Monte Carlo within the `Algorithms and Libraries for Physics Simulations' (ALPS) library~\cite{Bauer2011}. For these simulations, we calculate the scaled superfluid stiffness for different system sizes ($6\times6,\,8\times8,\,\mathrm{and}\,10\times10$) and employed the Binder cumulant to extrapolate the critical point of the phase transition~\cite{Bauer2011}.  After verifying convergence for the number of sweeps and thermalization steps, the number of Monte Carlo sweeps was set to 90000 and the number of Monte Carlo steps after thermalization was set to $9.6\times10^6$. 
At the tip of the $n = 1$ Mott lobe, the extrapolated phase transition point at $T=0$ is  $(U/J)_\mathrm{_c}\approx25.5-26$. 
For the frustrated system, in contrast, QMC cannot make any faithful prediction on the critical value for the superfluid-Mott insulator transition. 

Note that the expected transition point depends not only on the interaction but also on the chemical potential $\mu$. In an inhomogeneous trapped system, the phase transition hence does not happen everywhere at the same critical $U/J$~\cite{Jaksch_PRL_1998}. A full simulation of the inhomogeneous system would be possible using QMC~\cite{Trotzky2010} in the unfrustrated case but would go beyond the scope of this work. We instead use the tip of the Mott Lobe since we start with a relatively homogeneous $n=1$ Mott insulator.

\begin{figure*}[t] 
    \centering
    \includegraphics[width=1\textwidth]{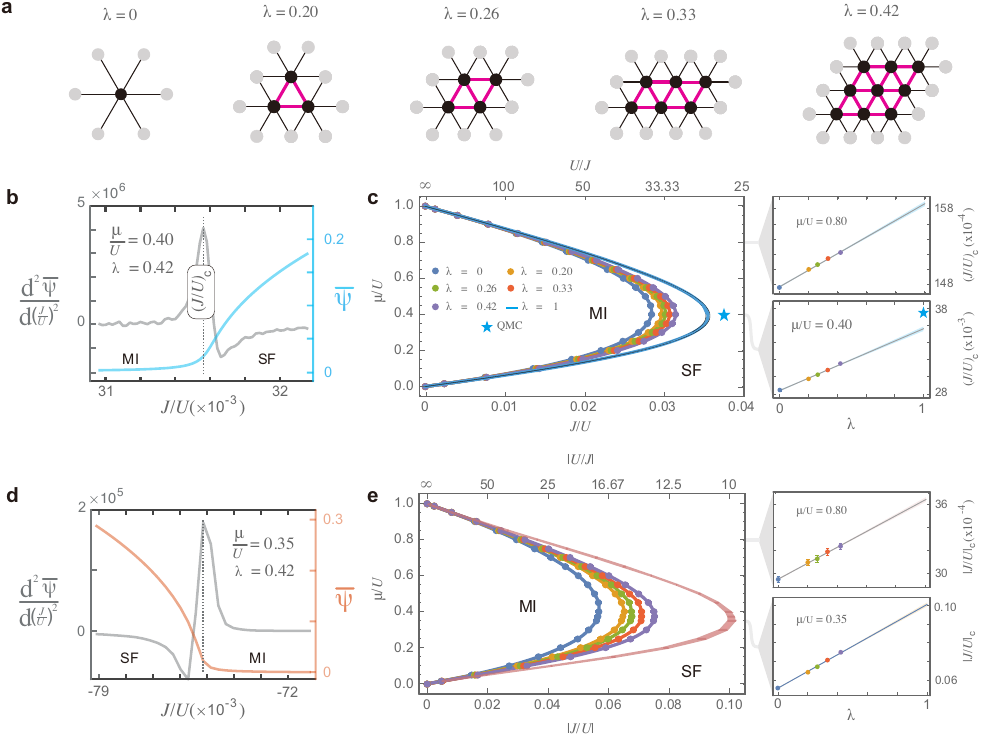}  
    \caption{\textbf{Cluster mean-field phase diagram calculation.} \textbf{a}, Triangular lattices with finite cluster size. Cluster sites are shown by black dots while sites outside the cluster are shown by grey dots. The number of bonds inside the cluster $N_\mathrm{c}$ shown by the thick purple lines and the number of bonds from the cluster to the mean-field $N_{\partial\mathrm{c}}$, quantify the finite lattice via the parameter $\lambda = \frac{N_\mathrm{c}}{N_\mathrm{c}+N_{\partial\mathrm{c}}}$. For an infinite lattice, $\lambda\to1$. \textbf{b}, For a $3\times3$ finite lattice with $\lambda = 0.42$ and $\mu/U = 0.40$, the order parameter $\Bar{\psi}$ for an \ufrs system (blue line), and its second derivative (grey line) are shown as a function of the normalized hopping $J/U$. The maximum value of the second derivative is attributed to the superfluid-Mott insulator phase transition point. \textbf{c}, Resulting phase diagram for $n=1$ Mott lobe for the \ufrs system for various system sizes. The blue band shows the $\pm\sigma$ confidence interval for the phase transition points extrapolated to an infinite lattice. The two outsets on the right show the method to extract the phase transition point for an infinite lattice (i.e., $\lambda = 1$ via linear extrapolation) from the calculations of the finite-sized lattice, at two different values of normalized chemical potential $\mu/U$. The blue star represents the prediction from quantum Monte-Carlo for the superfluid-Mott insulator transition. 
    \textbf{d},\textbf{e}, Equivalent plots for the \frs case.  
    } \label{fig:cMFTDetail} 
\end{figure*}  
\subsubsection{Cluster mean-field} \label{subsec:cMFTDiscussion}

Since quantum Monte Carlo cannot faithfully estimate the transition point in the frustrated case, we follow the calculation in~\cite{Yamamoto2020} and perform a cluster mean-field calculation for various lattice sizes shown by the black dots in Fig.~\ref{fig:cMFTDetail}\textbf{a}. Sites at the boundary of the cluster are connected to the mean-field via the thin black lines. The parameter $\lambda$ quantifies the finite-size cluster via $\lambda = \frac{N_\mathrm{c}}{N_\mathrm{c}+N_{\partial_\mathrm{c}}}$, where $N_\mathrm{c}$ is the number of bonds inside the cluster (purple lines), and $N_{\partial_\mathrm{c}}$ is the number of bonds from the cluster to the mean-field (thin black lines). For a given cluster, we estimate the transition point from the maximum of the second derivative (shown by the grey lines in Fig.~\ref{fig:cMFTDetail}\textbf{b \& d}) of the order parameter $\Bar{\psi}$ (shown by the light blue and red lines in Fig.~\ref{fig:cMFTDetail}\textbf{b \& d}, respectively). We then linearly extrapolate to the infinite lattice (i.e., $\lambda = 1$), see Fig.~\ref{fig:cMFTDetail}\textbf{c \& e}. For the \ufrs system we consider a superfluid at the $\mathbf{\Gamma}$ point, and for the \frs system, we use a superfluid at the $\mathbf{K}$ point~\cite{Yamamoto2020}. 

For our calculation, we have chosen a maximum boson occupation of three for each site, and the convergence tolerance during the self-consistent loop was set to $10^{-4}-10^{-5}$. 

\subsection{Critical Exponents}
For the \ufrs case, the MI-SF transition is in the 3DXY universality class with $z=1$~\cite{Fisher1989}. Together with the exponent $\nu\approx 0.67$~\cite{Wang2017} from the 3DXY model, this results in:
\[b_\mathrm{ufs.}=\dfrac{\nu}{1+\nu z}\approx 0.4\] 

In the sequential transition scenario in the \frs case, the transition from $\chi$-MI to $\chi$-SF would be of the same 3DXY type as the \ufrs case above. However, the preceding chirality-breaking transition from MI to $\chi$-MI would instead be expected to be in the 2D Ising universality class with $\nu = 1$~\cite{Kardar2007} and z = 2.14(2)~\cite{Adzhemyan2022}, resulting in:
\[b_\mathrm{frs.}=\dfrac{\nu}{1+\nu z}\approx 0.32\]

\section*{Acknowledgements}
We thank Liam Crane, Damien Bloch, and Tim Rein for their contributions during the building of the experimental setup and assistance with parts of the measurements. We would like to express our gratitude to Ehud Altman, Frédéric Chevy, Nigel Cooper, Ippei Danshita, Andr\'e Eckardt, Andreas Läuchli, Ludwig Mathey, Sebastiano Peotta, and Päivi Törmä for stimulating discussions. This work was supported by the European Commission ERC Starting Grant QUASICRYSTAL, the EPSRC Frontier Grant KAGOME (EP/X032795/1), the EPSRC Programme Grants DesOEQ (EP/P009565/1) and QQQS (EP/Y01510X/1), and the UK Quantum Technology Hub QCS (EP/T001082/1).



\bibliography{references}

\end{document}